\documentclass[a4paper,11pt]{article}
\usepackage{pos}
\usepackage{amsmath}
\usepackage{color} 

\newcommand\MSbar{$\overline{\rm MS}$}

\title{Testing nucleation calculations for strong phase transitions}

\author[a]{Oliver Gould}
\author[b]{Anna Kormu}
\author*[b]{David J. Weir}

\affiliation[a]{School of Physics and Astronomy,
University of Nottingham,
Nottingham NG7 2RD,
United Kingdom}

\affiliation[b]{Department of Physics and Helsinki Institute of Physics, P.O.~Box 64,
FI-00014 University of Helsinki, Finland}

\emailAdd{oliver.gould@nottingham.ac.uk}
\emailAdd{anna.kormu@helsinki.fi}
\emailAdd{david.weir@helsinki.fi}

\abstract{Accurate calculations of the nucleation rate $\Gamma$ for first order phase transitions are important for determining their observable consequences in particle physics and cosmology. Perturbative calculations are often used, but they are incomplete and should be tested against fully nonperturbative lattice simulations. We simulate nucleation on the lattice in a scalar field theory with a tree-level barrier, a scenario which should be well described by perturbation theory. Our computation of the nucleation rate, however, only shows qualitative agreement with the perturbative result. This motivates further study of nucleation on the lattice and to higher orders in perturbation theory.

\bigskip
{\small \textit{Preprint:  HIP-2025-7/TH}}
}

\FullConference{The 41st International Symposium on Lattice Field Theory (LATTICE2024)\\
 28 July - 3 August 2024\\
Liverpool, UK\\}

\renewcommand{\logo}{\relax}

\graphicspath{
  {./}
  {./figures/}
}

\begin{document}
\maketitle

\section{Introduction}
As the universe expanded and cooled, it is likely to have experienced one or more phase transitions. In a first-order phase transition, below a critical temperature $T_\mathrm{c}$ there is a discontinuous jump of some order parameter; the transition would consist of the nucleation, expansion and merging of bubbles of the new phase.

In the minimal Standard Model there is, rather than phase transition, a crossover -- but many theories of physics beyond the Standard Model predict a first-order phase transition. Such a phase transition can help to resolve the problem of the matter-antimatter asymmetry of the universe, or explain the observed abundance of dark matter. Moreover, if the phase transition were strong enough, it could produce an observable gravitational wave signal from the collision of the bubbles and possible inhomogeneities set up in the primordial plasma. For a first-order phase transition at around the electroweak scale, the resulting signal could be detected by the future space-based gravitational wave detector LISA, if it were sufficiently strong~\cite{LISACosmologyWorkingGroup:2022jok}.

The rate of bubble nucleation is a key quantity characterising the dynamics of first-order phase transitions. In field theory, the first complete computation of the rate was carried out by Langer~\cite{Langer:1967ax, Langer:1969bc}, whose method was based on a saddlepoint approximation to a path integral. This approach was later generalised and adapted to relativistic and quantum theories theories by Coleman~\cite{Coleman:1977py}, Affleck~\cite{Affleck:1980ac} and Linde~\cite{Linde:1981zj}. Yet these early works gave different expressions for the nucleation rate in the high temperature regime, a discrepancy which remains unresolved.

First-order phase transitions are also of interest in condensed matter systems. Experiments testing nucleation have been performed for systems including ferromagnetic superfluids~\cite{Zenesini:2023afv} and the A-B transition in $\,^3\mathrm{He}$~\cite{QUEST-DMC:2024crp}. The results for ferromagnetic superfluids are in good agreement with theory, whereas for $\,^3\mathrm{He}$ there is a longstanding discrepancy.

Lattice simulations provide an alternative way to probe the validity of nucleation theory, and may help to resolve some of the puzzles in both theoretical and experimental work. Studies involving observing the nucleation of bubbles in real-time simulations were carried out in Refs.~\cite{Alford:1993zf,Alford:1993ph,Borsanyi:2000ua,Batini:2023zpi,Pirvu:2024nbe}, but a more efficient method was developed by Moore and Rummukainen in Ref.~\cite{Moore:2000jw}. This uses multicanonical simulations to generate the highly-suppressed critical bubble configurations rather than waiting for them to appear. In this and subsequent work in Ref.~\cite{Moore:2001vf}, the motivation was the radiatively-induced phase transition in the minimal Standard Model. Nucleation in the minimal Standard Model was studied again in Ref~\cite{Gould:2022ran}, under the assumption that any new physics would be heavy enough to integrate out. However, with the recent interest in strong first-order phase transitions in models with a tree-level barrier, it is worth revisiting the simulational approaches to computing the nucleation rate. We expect perturbative calculations to perform better in scenarios with a stronger, tree-level barrier than where the barrier is radiatively induced, based on experiences with quantities such as the discontinuity in the field condensate~\cite{Gould:2021dzl}.

\subsection{Our model}

We study a toy model consisting of a single real scalar field with a tree-level potential barrier. We work in the high-temperature limit and assume that the physics of the phase transition -- and bubble nucleation in particular -- is well described by the long-wavelength modes and thus consider a dimensionally-reduced three-dimensional effective theory
\begin{equation}
  \mathscr{L}_\text{eff}  = \frac{1}{2}\partial_i \phi \partial_i \phi + V_3(\phi), \quad
  \text{where} \quad V_3(\phi) = \sigma_3 \phi + \frac{m_3^2}{2} \phi^2 + \frac{g_3}{3!} \phi^3 + \frac{\lambda_3}{4!} \phi^4,
\end{equation}
with spatial index $i \in \{1,2,3\}$.
  By matching observables, the parameters $\sigma_3$, $m_3^2$, $g_3$, and $\lambda_3$ can be expressed in terms of the original four-dimensional Lagrangian parameters and the temperature $T$.

The
thermodynamics of the phase transition in this model was carefully studied in Ref.~\cite{Gould:2021dzl}. The `scalnuc' code used for that paper was extended to include realtime simulations for the present work~\cite{ScalnucRelease}. Our lattice action is the $O(a^2)$ improved version of what was used for that study,
\begin{equation}
  S_\text{lat} = \sum_x a^3 \bigg[
  - \frac{1}{2} Z_\phi \phi_x (\nabla_\text{lat}^2 \phi)_x
  + \sigma_\text{lat} \phi_x
  + \frac{1}{2} Z_\phi Z_m m^2_\text{lat} \phi_x^2
  + \frac{1}{4!} Z_\phi^2 \lambda_\text{lat} \phi_x^4
  \bigg],
  \label{eq:lattice_action}
\end{equation}
where $a$ is the lattice spacing, and the cubic term has been absorbed into a constant field shift.
The $O(a^2)$ improvement determines the relation between $\kappa_\text{lat}$ and their three-dimensional continuum equivalents in the {\MSbar} scheme, $\kappa_\text{\MSbar}$, through exact lattice-continuum relations $\kappa_\text{lat} = \kappa_\text{\MSbar} + \delta\kappa$ as well as multiplicative constants $Z_\phi$, $Z_m$. For details, please see the Supplemental Material of Ref.~\cite{Gould:2024chm}.

\section{Simulations}

In order to compute the nucleation rate, we first simulate the theory~(\ref{eq:lattice_action}) using lattice Monte Carlo methods. Because the tunnelling between the metastable and stable phases is highly suppressed, we use the multicanonical method to overcome the barrier~\cite{Berg:1992qua}. In short, this consists of adding a weight function $W$ to the lattice action that encourages tunnelling, and which can be reweighted away to produce results for the original action. One must first generate $W$, which usually requires an additional simulation.

We select an `order parameter' $\theta_\text{op}$ that distinguishes between the field values in the two phases, and use this to construct a multicanonical weight function $W[\theta_\text{op}]$. For our theory, a simple volume average of the field value $\theta_\text{op, lin} = \overline{\phi}$ would seem, at first glance, to be sufficient.

Once the weight function has been successfully generated, we can simulate the resulting theory with lattice action $S_\text{lat}+W$ to obtain configurations that lie in the highly suppressed region between the two phases. These mixed-phase configurations can take the form of bubbles, slabs, or cylinders. If the box is sufficiently large, then the local maximum of the free energy between the two phases corresponds to a bubble. Furthermore, we describe the set of field configurations corresponding to this maximum the `separatrix' between the two phases (sometimes also referred to as the `transition surface'). In order for us to be working with critical bubbles rather than slabs or cylinders which are subject to finite-volume effects, our lattices must be sufficiently large. For this reason, we also use cubic lattices as there is no advantage in the current study to any other geometry. These geometrical issues are discussed further in Refs.~\cite{Rummukainen:2025pjj,Moore:2000jw,Moore:2001vf}.

Ensuring that the maximum of the free energy is a bubble motivates larger lattices, but this in turn leads to a new problem. In addition to mixed-phase fluctuations, there are also bulk fluctuations -- fluctuations around the metastable and stable phase. Since the critical bubble has a fixed, finite size, as the volume of the lattice is increased, configurations near each vacuum come to be dominated by the bulk phase fluctuations; this can overwhelm the critical bubble configurations 
 -- and 
it becomes difficult to determine whether a given configuration with a near-critical value of the order parameter consists of bulk fluctuations or is a true critical bubble (an issue alluded to in Ref.~\cite{Gould:2022ran}).

One therefore cannot continue to arbitrarily large lattices with this method. However, the extent of the bulk fluctuations depends on the exact choice of order parameter and we can alleviate the symptoms by using the modified order parameter $\theta_\text{op} = \overline{\phi^2}-2A\overline{\phi}$, where $A$ is an arbitrary constant. If $A$ is selected to be close to the metastable vacuum peak value of $\overline{\phi}$, then we found that this order parameter dramatically suppressed bulk phase fluctuations. For a more detailed discussion of alternative order parameters and pseudo-order parameters for studying critical bubble configurations, see Ref.~\cite{Rummukainen:2025pjj} in these proceedings. In particular, the `order parameter' does not even need to be a true order parameter, so long as the critical bubble configurations can be distinguished from the bulk phases.

\begin{figure}
  \centering
  \includegraphics[width=0.48\textwidth]{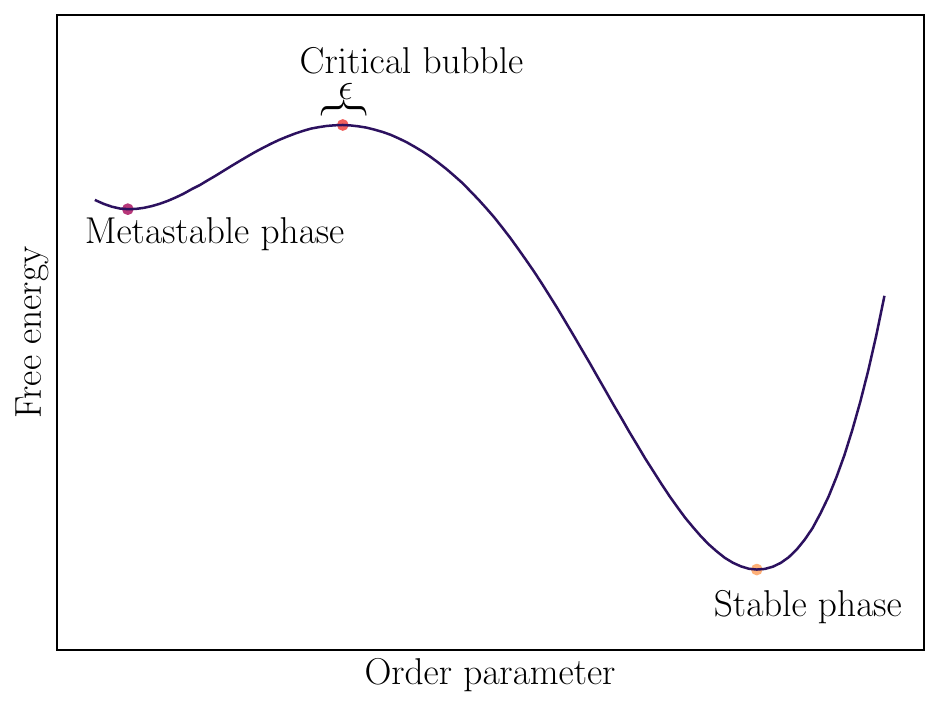}
  \caption{Free energy as the function of the order parameter at some temperature below $T_c$.
  The metastable and stable phases are separated by an exponentially suppressed mixed phase. The maximum of the free energy in this mixed phase corresponds to the critical bubble. Configurations near the critical bubble - the separatrix - are drawn from a narrow range $\epsilon$ and act as initial conditions for the real time dynamics of the system. }
  \label{fig:hist}
\end{figure}

Once we are able to resolve the configurations corresponding to critical bubbles, we can compute the probability density of finding a near-critical bubble relative to the metastable phase,
\begin{equation}
  P_c^{\mathrm{normalised}}=\frac{P(|\theta_\text{op}-\theta_\mathrm{c}|<\epsilon/2)}{\epsilon P(\theta_\text{op}<\theta_\mathrm{c})},
  \label{eq:critical_bubble_probability}
\end{equation}
where $\theta_\mathrm{c}$ is the order parameter value corresponding to the critical bubble (and hence maximum of the free energy), and $\epsilon$ is a small constant (see Fig.~\ref{fig:hist}). Note that this critical bubble probability density depends on the choice of $\theta_\text{op}$ and $\epsilon$, as demonstrated in the results of Ref.~\cite{Rummukainen:2025pjj}.

\subsection{Computing the nucleation rate}

The critical bubble probability above is obtained with standard multicanonical lattice simulations. However, we use the same configurations for which $\theta_\text{op} \in [\theta_\mathrm{c} - \frac{\epsilon}{2}, \theta_\mathrm{c} + \frac{\epsilon}{2}]$ 
as the starting point for real-time simulations. We then determine the fraction of these configurations which tunnel and thus correspond to true critical bubbles, by evolving each configuration both forwards and backwards in the presence of a thermal bath, to see which phase the system travels towards. In this way, the exact choice of $\epsilon$ is not important.

There are three ingredients to the overall nucleation rate: the probability of obtaining a near-critical bubble configuration, the flux through the separatrix surface in field space, and the fraction of near-critical configurations that actually tunnel. Assuming that the flux depends principally upon physics on short timescales whereas the probability of a configuration tunnelling depends on physics on much longer timescales~\cite{Moore:2001vf}
, the overall rate $\Gamma$ in a volume $\mathcal{V}$ factorises:
\begin{equation}
  \Gamma \mathcal{V} \approx P_c^{\mathrm{normalised}}\frac{1}{2} \left\langle \mathrm{flux}\right\rangle \left\langle \bf{d}\right\rangle,
\end{equation}
where $\langle\mathrm{flux}\rangle = \left< \left| \frac{\Delta \theta_\mathrm{op}}{\Delta t} \right|_{\theta_\mathrm{c}} \right>
   = \sqrt{\frac{8}{\pi \mathcal{V}}(\theta_\mathrm{c}+A^2)}$ for our choice of order parameter $\theta_\mathrm{op}$, assuming that the momentum field is Gaussian. The quantity $\mathbf{d}$ is related\footnote{See Ref.~\cite{Moore:2000jw}, Appendix A for a justification of this approach.} to the number of configurations that tunnel during realtime simulations, normalised to the number of times $N_{\mathrm{crossings}}$ a tunnelling trajectory crosses $\theta_\mathrm{c}$:
   \begin{equation}
    \mathbf{d} = 
  \frac{\delta_{\mathrm{tunnel}}}{N_{\mathrm{crossings}}}, \quad \text{where} \quad \delta_{\mathrm{tunnel}} = \begin{cases}
  1 & \text{if trajectory tunnels} \\
  0 & \text{if trajectory does not tunnel}
  \end{cases}
  \end{equation}
      -- the factor of $\frac{1}{2}$ is to account for the fact we are only interested in nucleation processes in one direction (i.e. from metastable to stable).

We use the fourth order accurate symplectic Forest-Ruth algorithm with momentum refresh to reproduce the effects of the critical bubbles being exposed to the thermal fluctuations of the primordial heat bath. This evolution algorithm is built out of a standard leapfrog evolution with timestep $\Delta t$:
\begin{align}
  \pi_{t+\frac{1}{2},x} &=\pi_{t,x} - \frac{1}{a^3}\frac{\partial H_\text{eff}}{\partial \phi_{t,x}}\frac{\Delta t}{2} ,\\
  \phi_{t+1,x} &= \phi_{t,x} + \pi_{t+\frac{1}{2},x}\Delta t ,\\
  \pi_{t+1,x} &=\pi_{t+\frac{1}{2},x} - \frac{1}{a^3}\frac{\partial H_\text{eff}}{\partial \phi_{t+1,x}}\frac{\Delta t}{2},
\end{align}
where $H_\text{eff}$ is the Hamiltonian for the system, corresponding to the action~(\ref{eq:lattice_action}) and $\pi$ is the conjugate momentum to the field $\phi$. One Forest-Ruth step consists of three successive updates with different timesteps:
\begin{equation}
  \Delta t_1 = (2-2^{1/3})^{-1} \Delta t ,
  \quad
  \Delta t_2 = -2^{1/3}(2-2^{1/3})^{-1} \Delta t, 
  \quad
  \Delta t_3 = \Delta t_1 . 
\end{equation}
After these three sub-steps, the Forest-Ruth algorithm has advanced one $\Delta t$ timestep forwards. After each step we perform a momentum refresh,
\begin{align}
  \pi_{t+0,x} &= \sqrt{1-\vartheta^2}\pi_{t-0,x} + \vartheta \xi_{t,x} ,\\
  \vartheta^2 &= 1-\exp(-2\gamma\Delta t),
\end{align}
where $\gamma$ is a damping term and $\xi$ satisfies 
\begin{equation}
  \langle \xi_{t, \mathbf{x}} \xi_{t', \mathbf{x}'} \rangle = \frac{2\gamma}{a^3 \Delta t} \delta_{t}^{t'}\delta_{x}^{x'}.
\end{equation}
We found empirically that a high-order symplectic algorithm was needed so that the Monte Carlo and real time stages of the calculation would agree on the location of the separatrix in field configuration space. Performing the above evolution is equivalent to evolving a Langevin equation for the system with damping $\gamma$ and noise $\xi$. We take $\gamma=1/L$ where $L$ is the box size, so that $\gamma\to 0_+$ in the infinite volume limit. This is chosen in order to reproduce pure Hamiltonian evolution which captures the evolution of the quantum theory at leading order~\cite{Aarts:1997kp}, while minimising finite-volume artefacts from the lattice heating up as the bubble expands~\cite{Moore:2001vf}.

\begin{figure}
  \centering
  \includegraphics[width=0.48\textwidth]{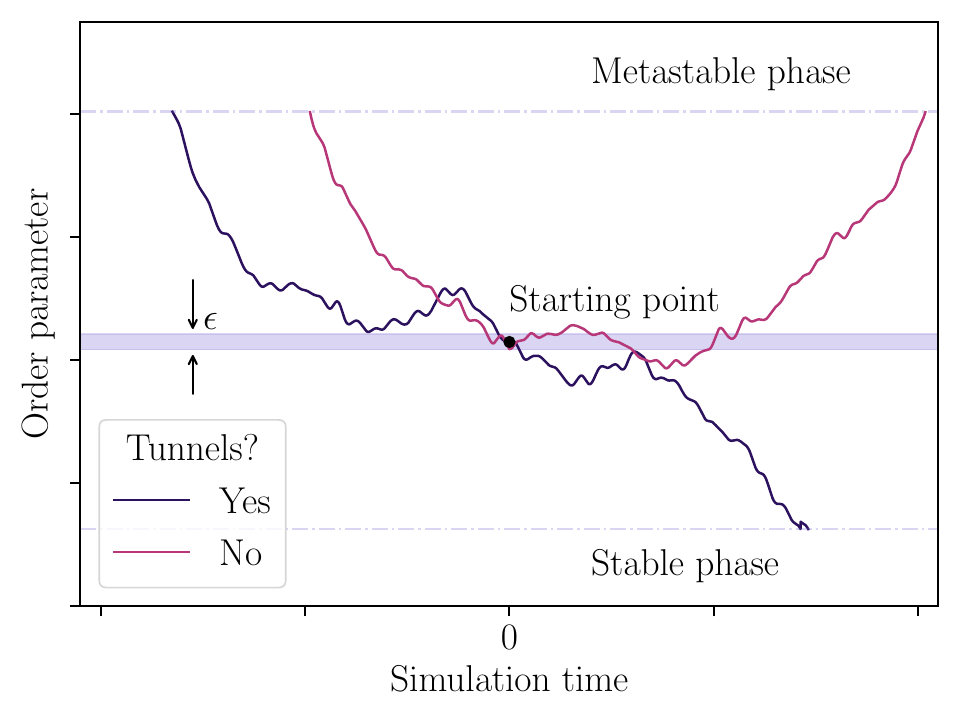}
  \caption{Schematic illustration of trajectories starting from
  near-critical bubble configurations, drawn from the narrow range $\epsilon$.
  Configurations are evolved backwards and forwards in time from the starting point and then combined into a full trajectory that is used to determine whether a given configuration tunnels or not. The system is evolved forwards and backwards for a given number of timesteps, or until the field leaves a predefined range (dashed lines) of the order parameter near $\theta_\mathrm{c}$, whichever happens first.}
  \label{fig:traj}
\end{figure}

Trajectories are evolved forward and backward from the same initial state (see Fig.~\ref{fig:traj}). The initial momenta are reversed for the backward evolution. 

\section{Results}
We carry out simulations at one benchmark parameter point, with $\mu_3 / \lambda_3 = 1$, $\sigma_3/\lambda_3^{5/2} = -0.016687$, $m_3^2/\lambda_3^2 = -0.082770$, and $g_3/\lambda_3^{3/2} = 0$.
For example, this corresponds to physics in the xSM theory of the Standard Model with a real scalar field with
$\lambda_3 = 144.23 \, \mathrm{GeV}$ and $T = 93.121 \, \mathrm{GeV}$.
For this example, the jump in the scalar condensate at the critical temperature is $\Delta \langle\bar{\phi}\rangle_\mathrm{c}/\sqrt{T_\mathrm{c}}\approx 1.67$.
We use jackknife error estimation for all our results.

\begin{figure}
  \centering
    \includegraphics[width=.48\textwidth]{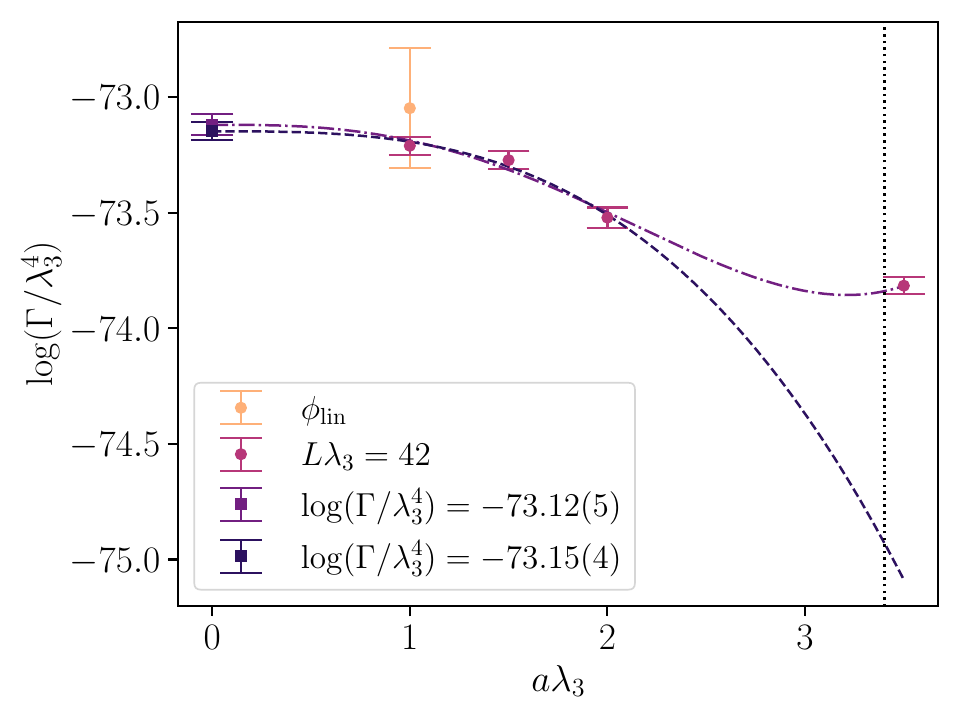}
  \hfill
    \includegraphics[width=.48\textwidth]{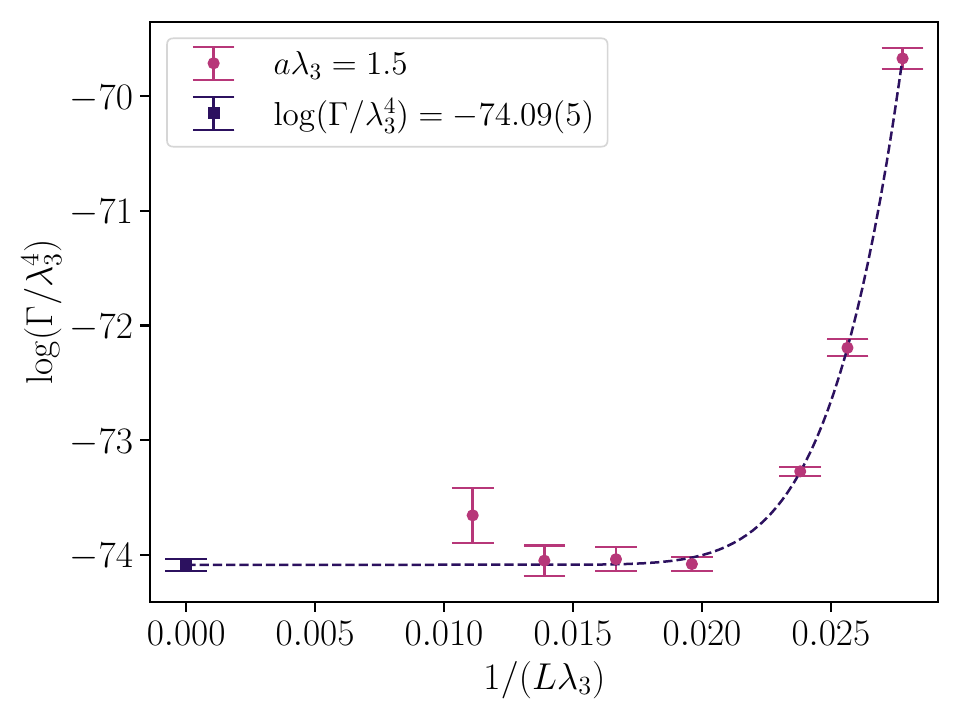}
  \caption{At left, the continuum extrapolation for fixed volume $L\lambda_3=42$, along with one point computed with the linear order parameter (denoted $\phi_\text{lin}$). Given our $O(a^2)$ improvement, we fit to cubic  $f(a) = b+ca^3$ and quartic $f(a) = b+ca^3+da^4$, noting also that our largest lattice spacing is comparable to the inverse screening mass; we thus exclude it from the cubic fit. At right, an extrapolation to infinite volume for $a\lambda_3=1.5$, with a fit $f(L) = b + c e^{-m_\mathrm{s}L}$, where $m_\mathrm{s}$ is expected to be the screening mass.
  Figures from Ref.~\cite{Gould:2024chm}. \label{fig:limits}}
\end{figure}

In Fig.~\ref{fig:limits}, we show the fitting and extrapolation to continuum and infinite volume limits. For one choice of lattice spacing ($a\lambda_3=1$) and volume ($L \lambda_3 = 42$), we also show the result of performing a comparable study but with the naive order parameter $\phi_\text{op, lin} \equiv \overline{\phi}$. We note that this results in considerably larger uncertainty than our final choice of order parameter; the principal source of this larger error is the difficulty in distinguishing the bubble configuration peak of the free energy from bulk fluctuations. To extrapolate to the infinite volume limit we fix $a\lambda_3=1.5$, based on our observation that lattice artefacts and statistical uncertainties are comparable at this lattice spacing.

\begin{figure}
  \centering
  \includegraphics[width=0.48\textwidth]{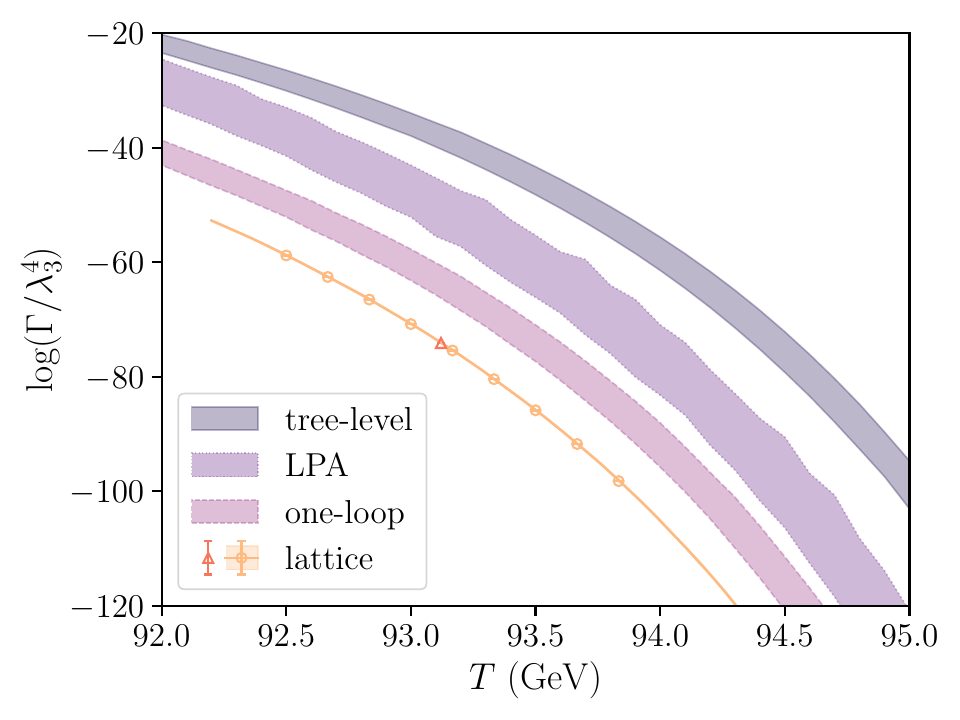}
  \caption{The nucleation rate as a function of temperature for the tree-level result, the local potential approximation (LPA), and one-loop perturbative results. For the tree-level and one-loop results the uncertainty bands are given by varying the renormalisation scale, while for the LPA result the uncertainty band depends on different choices for handling the complex potential. The continuum-extrapolated lattice nucleation result based directly on our simulations is shown as an orange triangle,
  while the orange circles are continuum-extrapolated, reweighted points. Finally, a continuously reweighted result for $a\lambda_3=1.5$, $L\lambda_3=60$ is given as an orange line.
Figure from Ref.~\cite{Gould:2024chm}.
\label{fig:reweighted}}
\end{figure}

Finally, in Fig.~\ref{fig:reweighted} we reweight our result to different temperatures and compare our findings to various analytical results. Varying the renormalisation scale yields an estimate of the uncertainty in the tree-level and perturbative results, while the local potential approximation (LPA) depends on how the imaginary parts of the potential are handled. The lattice value for the nucleation rate has far smaller statistical errors than any of these, both at our continuum-extrapolated point and the constant-volume reweighted line. Furthermore, our result is also considerably lower than the one-loop result throughout the reweightable range.  

\section{Discussion}
Our results show a discrepancy between the lattice and one-loop calculations.
The parameter point studied was chosen to be well described by perturbation theory. For the latent heat, the discrepancy between the lattice and one-loop results is less than 1\%, and higher loop results agree within statistical errors~\cite{Gould:2021dzl}. However, this is not the case for bubble nucleation where we find that the disagreement in $|\log \Gamma |$ is 20\% at one-loop and 100\% at tree-level. 

Although it is possible the results can be reconciled with a two-loop perturbative calculation, what if the discrepancy cannot be explained by going to the next loop order?
There are other possible explanations -- beyond the need for the next loop-order -- as to why we might be seeing this large difference. It has been suggested that there may well exist other relevant saddlepoints besides the critical bubble~\cite{Andreassen:2016cvx}, or that the saddlepoint approximation breaks down~\cite{Ekstedt:2022tqk}.

Further nucleation studies using the paradigm developed by Moore and Rummukainen (and applied here) would also be welcome. We have worked with just one benchmark point in one particular theory. However, it is a somewhat involved framework, and there are technical challenges around (for example) handling bulk phase fluctuations that we have only started to address.

One can also directly simulate nucleation with the Langevin equation, a technique which has seen a revival of interest~\cite{Batini:2023zpi, Pirvu:2023plk}. These simulations are not without their own challenges, however~\cite{Pirvu:2024ova}, and by their very nature cannot in their current form reach the very low rates of nucleation studied in the present work.

For phase transitions in the early universe, the systematic theoretical uncertainty in bubble nucleation rate calculations is one of the biggest contributors to uncertainty in the amplitude of the resulting gravitational wave power spectrum~\cite{Gould:2021oba}. Reducing this uncertainty is crucial if one is to predict or constrain physics beyond the Standard Model through the gravitational wave power spectrum resulting from a phase transition in such models.

\acknowledgments

O.G. (ORCID ID 0000-0002-7815-3379) was
  supported by
  the Research Funds of the University of Helsinki,
  U.K. Science and Technology Facilities Council (STFC) Consolidated Grant
  ST/T000732/1,
  a Research Leadership Award from the Leverhulme Trust,
  and a Royal Society Dorothy Hodgkin Fellowship.
  A.K. (ORCID ID 0000-0002-0309-3471) was supported by Research Council of Finland grant no. 328958 and a travel grant from the Jenny and Antti Wihuri Foundation.
  D.J.W. (ORCID ID 0000-0001-6986-0517) was
  supported by Research Council of Finland grant nos. 324882, 328958, 349865 and 353131.
  The authors also wish to acknowledge CSC -- IT Center for Science, Finland, for computational resources.

\section*{Data access statement}
Data supporting this manuscript is available from Zenodo~\cite{gould_2024_11085693}, and the source code is archived at Ref.~\cite{ScalnucRelease}.

\bibliographystyle{JHEP}
\bibliography{nucleation_rate}

\newcommand{\noop}[1]{}

\providecommand{\href}[2]{#2}\begingroup\raggedright\begin{thebibliography}{10}

\bibitem{LISACosmologyWorkingGroup:2022jok}
{\scshape LISA Cosmology Working Group} collaboration, \emph{{Cosmology with the Laser Interferometer Space Antenna}}, \href{https://doi.org/10.1007/s41114-023-00045-2}{\emph{Living Rev. Rel.} {\bfseries 26} (2023) 5} [\href{https://arxiv.org/abs/2204.05434}{{\ttfamily 2204.05434}}].

\bibitem{Langer:1967ax}
J.S.~Langer, \emph{{Theory of the condensation point}}, \href{https://doi.org/10.1016/0003-4916(67)90200-X}{\emph{Annals Phys.} {\bfseries 41} (1967) 108}.

\bibitem{Langer:1969bc}
J.S.~Langer, \emph{{Statistical theory of the decay of metastable states}}, \href{https://doi.org/10.1016/0003-4916(69)90153-5}{\emph{Annals Phys.} {\bfseries 54} (1969) 258}.

\bibitem{Coleman:1977py}
S.R.~Coleman, \emph{{The Fate of the False Vacuum. 1. Semiclassical Theory}}, \href{https://doi.org/10.1103/PhysRevD.16.1248}{\emph{Phys. Rev. D} {\bfseries 15} (1977) 2929}.

\bibitem{Affleck:1980ac}
I.~Affleck, \emph{{Quantum Statistical Metastability}}, \href{https://doi.org/10.1103/PhysRevLett.46.388}{\emph{Phys. Rev. Lett.} {\bfseries 46} (1981) 388}.

\bibitem{Linde:1981zj}
A.D.~Linde, \emph{{Decay of the False Vacuum at Finite Temperature}}, \href{https://doi.org/10.1016/0550-3213(83)90072-X}{\emph{Nucl. Phys. B} {\bfseries 216} (1983) 421}.

\bibitem{Zenesini:2023afv}
A.~Zenesini, A.~Berti, R.~Cominotti, C.~Rogora, I.G.~Moss, T.P.~Billam et~al., \emph{{False vacuum decay via bubble formation in ferromagnetic superfluids}}, \href{https://doi.org/10.1038/s41567-023-02345-4}{\emph{Nature Phys.} {\bfseries 20} (2024) 558} [\href{https://arxiv.org/abs/2305.05225}{{\ttfamily 2305.05225}}].

\bibitem{QUEST-DMC:2024crp}
{\scshape QUEST-DMC} collaboration, \emph{{A-B Transition in Superfluid $^3$He and Cosmological Phase Transitions}}, \href{https://doi.org/10.1007/s10909-024-03151-9}{\emph{J. Low Temp. Phys.} {\bfseries 215} (2024) 495} [\href{https://arxiv.org/abs/2401.07878}{{\ttfamily 2401.07878}}].

\bibitem{Alford:1993zf}
M.G.~Alford, H.~Feldman and M.~Gleiser, \emph{{Thermal activation of metastable decay: Testing nucleation theory}}, \href{https://doi.org/10.1103/PhysRevD.47.R2168}{\emph{Phys. Rev. D} {\bfseries 47} (1993) R2168}.

\bibitem{Alford:1993ph}
M.G.~Alford and M.~Gleiser, \emph{{Metastability in two-dimensions and the effective potential}}, \href{https://doi.org/10.1103/PhysRevD.48.2838}{\emph{Phys. Rev. D} {\bfseries 48} (1993) 2838} [\href{https://arxiv.org/abs/hep-ph/9304245}{{\ttfamily hep-ph/9304245}}].

\bibitem{Borsanyi:2000ua}
S.~Borsanyi, A.~Patkos, J.~Polonyi and Z.~Szep, \emph{{Fate of the classical false vacuum}}, \href{https://doi.org/10.1103/PhysRevD.62.085013}{\emph{Phys. Rev. D} {\bfseries 62} (2000) 085013} [\href{https://arxiv.org/abs/hep-th/0004059}{{\ttfamily hep-th/0004059}}].

\bibitem{Batini:2023zpi}
L.~Batini, A.~Chatrchyan and J.~Berges, \emph{{Real-time dynamics of false vacuum decay}}, \href{https://doi.org/10.1103/PhysRevD.109.023502}{\emph{Phys. Rev. D} {\bfseries 109} (2024) 023502} [\href{https://arxiv.org/abs/2310.04206}{{\ttfamily 2310.04206}}].

\bibitem{Pirvu:2024nbe}
D.~P\^\i{}rvu, A.~Shkerin and S.~Sibiryakov, \emph{{Thermal false vacuum decay in (1+1) dimensions: Evidence for nonequilibrium dynamics}}, \href{https://doi.org/10.1142/S0217751X24450076}{\emph{Int. J. Mod. Phys. A} {\bfseries 39} (2024) 2445007} [\href{https://arxiv.org/abs/2408.06411}{{\ttfamily 2408.06411}}].

\bibitem{Moore:2000jw}
G.D.~Moore and K.~Rummukainen, \emph{{Electroweak bubble nucleation, nonperturbatively}}, \href{https://doi.org/10.1103/PhysRevD.63.045002}{\emph{Phys. Rev. D} {\bfseries 63} (2001) 045002} [\href{https://arxiv.org/abs/hep-ph/0009132}{{\ttfamily hep-ph/0009132}}].

\bibitem{Moore:2001vf}
G.D.~Moore, K.~Rummukainen and A.~Tranberg, \emph{{Nonperturbative computation of the bubble nucleation rate in the cubic anisotropy model}}, \href{https://doi.org/10.1088/1126-6708/2001/04/017}{\emph{JHEP} {\bfseries 04} (2001) 017} [\href{https://arxiv.org/abs/hep-lat/0103036}{{\ttfamily hep-lat/0103036}}].

\bibitem{Gould:2022ran}
O.~Gould, S.~G\"uyer and K.~Rummukainen, \emph{{First-order electroweak phase transitions: A nonperturbative update}}, \href{https://doi.org/10.1103/PhysRevD.106.114507}{\emph{Phys. Rev. D} {\bfseries 106} (2022) 114507} [\href{https://arxiv.org/abs/2205.07238}{{\ttfamily 2205.07238}}].

\bibitem{Gould:2021dzl}
O.~Gould, \emph{{Real scalar phase transitions: a nonperturbative analysis}}, \href{https://doi.org/10.1007/JHEP04(2021)057}{\emph{JHEP} {\bfseries 04} (2021) 057} [\href{https://arxiv.org/abs/2101.05528}{{\ttfamily 2101.05528}}].

\bibitem{ScalnucRelease}
O.~Gould, A.~Kormu and D.J.~Weir, ``Scalnuc release 2.1.0.'' SWHID \href{https://archive.softwareheritage.org/swh:1:rel:94596986a4cf3bfd61ed75e34de3fe46a66d4753;origin=https://bitbucket.org/og113/scalnuc.git;visit=swh:1:snp:85d8b3bf7b4550676db3ca4913273cd89ea0a40f}{swh:1:rel:94596986a4cf3bfd61ed75e34de3fe46a66d4753; \\ origin=https://bitbucket.org/og113/scalnuc.git; \\ visit=swh:1:snp:85d8b3bf7b4550676db3ca4913273cd89ea0a40f}, 2025.

\bibitem{Gould:2024chm}
O.~Gould, A.~Kormu and D.J.~Weir, \emph{{Nonperturbative test of nucleation calculations for strong phase transitions}}, {\emph{Phys. Rev. D} (\noop{3002}2025, in press) } [\href{https://arxiv.org/abs/2404.01876}{{\ttfamily 2404.01876}}].

\bibitem{Berg:1992qua}
B.A.~Berg and T.~Neuhaus, \emph{{Multicanonical ensemble: A New approach to simulate first order phase transitions}}, \href{https://doi.org/10.1103/PhysRevLett.68.9}{\emph{Phys. Rev. Lett.} {\bfseries 68} (1992) 9} [\href{https://arxiv.org/abs/hep-lat/9202004}{{\ttfamily hep-lat/9202004}}].

\bibitem{Rummukainen:2025pjj}
K.~Rummukainen, R.~Sepp\"a and D.J.~Weir, \emph{{Resolving the critical bubble in $\mathrm{SU}(8)$ deconfinement transition}}, \href{https://doi.org/10.22323/1.466.0434}{\emph{PoS} {\bfseries LATTICE2024} (2025) 434} [\href{https://arxiv.org/abs/2501.17593}{{\ttfamily 2501.17593}}].

\bibitem{Aarts:1997kp}
G.~Aarts and J.~Smit, \emph{{Classical approximation for time dependent quantum field theory: Diagrammatic analysis for hot scalar fields}}, \href{https://doi.org/10.1016/S0550-3213(97)00723-2}{\emph{Nucl. Phys. B} {\bfseries 511} (1998) 451} [\href{https://arxiv.org/abs/hep-ph/9707342}{{\ttfamily hep-ph/9707342}}].

\bibitem{Andreassen:2016cvx}
A.~Andreassen, D.~Farhi, W.~Frost and M.D.~Schwartz, \emph{{Precision decay rate calculations in quantum field theory}}, \href{https://doi.org/10.1103/PhysRevD.95.085011}{\emph{Phys. Rev. D} {\bfseries 95} (2017) 085011} [\href{https://arxiv.org/abs/1604.06090}{{\ttfamily 1604.06090}}].

\bibitem{Ekstedt:2022tqk}
A.~Ekstedt, \emph{{Bubble nucleation to all orders}}, \href{https://doi.org/10.1007/JHEP08(2022)115}{\emph{JHEP} {\bfseries 08} (2022) 115} [\href{https://arxiv.org/abs/2201.07331}{{\ttfamily 2201.07331}}].

\bibitem{Pirvu:2023plk}
D.~P\^\i{}rvu, M.C.~Johnson and S.~Sibiryakov, \emph{{Bubble velocities and oscillon precursors in first-order phase transitions}}, \href{https://doi.org/10.1007/JHEP11(2024)064}{\emph{JHEP} {\bfseries 11} (2024) 064} [\href{https://arxiv.org/abs/2312.13364}{{\ttfamily 2312.13364}}].

\bibitem{Pirvu:2024ova}
D.~P\^\i{}rvu, A.~Shkerin and S.~Sibiryakov, \emph{{Thermal False Vacuum Decay Is Not What It Seems}},  \href{https://arxiv.org/abs/2407.06263}{{\ttfamily 2407.06263}}.

\bibitem{Gould:2021oba}
O.~Gould and T.V.I.~Tenkanen, \emph{{On the perturbative expansion at high temperature and implications for cosmological phase transitions}}, \href{https://doi.org/10.1007/JHEP06(2021)069}{\emph{JHEP} {\bfseries 06} (2021) 069} [\href{https://arxiv.org/abs/2104.04399}{{\ttfamily 2104.04399}}].

\bibitem{gould_2024_11085693}
O.~Gould, A.~Kormu and D.J.~Weir, ``Final data for paper \textsl{A nonperturbative test of nucleation calculations for strong phase transitions}.'' DOI \href{https://doi.org/10.5281/zenodo.10891523}{10.5281/zenodo.10891523}, 2024.

\end{thebibliography}\endgroup

\end{document}